\documentclass[%
 aip,
 amsmath,amssymb,
 reprint,%
]{revtex4-1}

\usepackage{graphicx}% Include figure files
\usepackage{dcolumn}% Align table columns on decimal point
\usepackage{bm}% bold math
\renewcommand{\vec}{\mathbf}
\newcommand{\PT}{\mathcal{P}\mathcal{T}}
\usepackage[utf8]{inputenc}
\usepackage[T1]{fontenc}
\usepackage{mathptmx}
\usepackage{etoolbox}
\usepackage[divpsnames]{xcolor}

\makeatletter
\def\@email#1#2{%
 \endgroup
 \patchcmd{\titleblock@produce}
  {\frontmatter@RRAPformat}
  {\frontmatter@RRAPformat{\produce@RRAP{*#1\href{mailto:#2}{#2}}}\frontmatter@RRAPformat}
  {}{}
}%
\makeatother

\begin{document}

% Use the \preprint command to place your local institutional report number 
% on the title page in preprint mode.
% Multiple \preprint commands are allowed.
%\preprint{}

\title{Non-Hermitian physics in magnetic systems }

% \email, \thanks, \homepage, \altaffiliation all apply to the current author.
% Explanatory text should go in the []'s, 
% actual e-mail address or url should go in the {}'s for \email and \homepage.
% Please use the appropriate macro for the type of information

% \affiliation command applies to all authors since the last \affiliation command. 
% The \affiliation command should follow the other information.

\author{Hilary M. Hurst}
\email[Corresponding Author: ]{hilary.hurst@sjsu.edu}
%\homepage[]{Your web page}
%\thanks{}
%\altaffiliation{}
\affiliation{Department of Physics \& Astronomy, San Jos\'{e} State University, San Jos\'{e}, California, 95192, USA}

\author{Benedetta Flebus}
%\email[]{flebus@bc.edu}
%\homepage[]{Your web page}
%\thanks{}
%\altaffiliation{}
\affiliation{Department of Physics, Boston College, 140 Commonwealth Avenue, Chestnut Hill, Massachusetts 02467, USA}

% Collaboration name, if desired (requires use of superscriptaddress option in \documentclass). 
% \noaffiliation is required (may also be used with the \author command).
%\collaboration{}
%\noaffiliation

\date{\today}

\begin{abstract}
%\textbf{I have noticed we do not have an abstract yet.}
Non-Hermitian Hamiltonians provide an alternative perspective on the dynamics of quantum and classical systems coupled non-conservatively to an environment. Once primarily an interest of mathematical physicists, the theory of non-Hermitian Hamiltonians has solidified and expanded to describe various physically observable phenomena in optical, photonic, and condensed matter systems. Self-consistent descriptions of quantum mechanics based on non-Hermitian Hamiltonians have been developed and continue to be refined. In particular, non-Hermitian frameworks to describe magnonic and hybrid magnonic systems have gained popularity and utility in recent years, with new insights into the magnon topology, transport properties, and phase transitions coming into view. Magnonic systems are in many ways a natural platform in which to realize non-Hermitian physics because they are always coupled to a surrounding environment and exhibit lossy dynamics. In this perspective we review recent progress in non-Hermitian frameworks to describe magnonic and hybrid magnonic systems, such as cavity magnonic systems and magnon-qubit coupling schemes. We discuss progress in understanding the dynamics of inherently lossy magnetic systems as well as systems with gain induced by externally applied spin currents. We enumerate phenomena observed in both purely magnonic and hybrid magnonic systems which can be understood through the lens of non-Hermitian physics, such as $\PT$ and Anti-$\PT$-symmetry breaking, dynamical magnetic phase transitions, non-Hermitian skin effect, and the realization of exceptional points and surfaces. Finally, we comment on some open problems in the field and discuss areas for further exploration. 
\end{abstract}

% insert suggested PACS numbers in braces on next line

\maketitle %\maketitle must follow title, authors, abstract and \pacs

% Body of paper goes here. Use proper sectioning commands. 
% References should be done using the \cite, \ref, and \label commands
\section{Introduction \label{Sec:Introduction}}
The postulates of quantum mechanics dictate that physical observables of quantum systems correspond to a Hamiltonian description of the system, where the Hamiltonian must be Hermitian. However, it has been known for some time that \emph{non-Hermitian} Hamiltonians can be quite useful in describing systems where energy or particle number is not conserved, despite the possible emergence of complex spectra~\cite{Bender1998, Bender1999, Bender2007, Heiss2012}. Non-Hermitian Hamiltonians have been widely used in the theory of exciton-polaritons~\cite{Gao2015, Comaron2020, Su2021}, photonics~\cite{Feng2017, el2019, Weidemann2020, Parto2021}, and other optical systems~\cite{El2018}. More recently, researchers have been developing non-Hermitian descriptions of magnonic systems, with an eye toward describing magnons in driven and dissipative spintronic systems. Through a non-Hermitian theoretical description, we can find new insights into both classical and quantum behavior in magnetic systems, and eventually develop and engineer new magnonic devices. 

Effective non-Hermitian Hamiltonians have been used to describe ordinary dissipation for some time (i.e. as an $H_{\rm eff}$ that would appear in any master equation). However, insights gained from the non-Hermitian description beyond this paradigm can help us to actually engineer and understand new phenomena in magnonic systems. Any non-Hermitian Hamiltonian is, strictly speaking, an approximation to more complicated dissipative dynamics which can be more comprehensively described using  nonlinear dynamics (classical case) or the master equation (in the quantum case). The non-Hermitian description cannot capture all processes that occur in open systems, such as probabilistic quantum jump processes. However, it remains a useful theoretical tool due to its simplicity and ability to reveal certain global properties of open systems through simple symmetry analysis and analysis of system properties around exceptional points (EPs) in parameter space, i.e. regions where eigenvectors coalesce.  Non-Hermitian Hamiltonians are particularly useful in Markovian systems that can be engineered to have a dynamical fixed point, for example in a system with balanced gain and loss, where it is known that, despite the non-Hermiticity of the system, quantum states can remain pure throughout their evolution~\cite{Brody2012}. 

Modeling systems through a non-Hermitian framework has already lead to a number of exciting theoretical and experimental results. Non-Hermitian systems can have markedly different topological properties from their Hermitian counterparts~\cite{Gong2018, Kawabata2019, Kawabata2020, Bergholtz2021}, leading to new symmetry classes and edge state properties, including edge mode `lasing'. This insight has lead to the experimental development of topological lasers in photonic systems~\cite{Bandres2018, Harari2018, Ezawa2022}. Encircling an exceptional point in a non-Hermitian system can lead to new dynamical behavior such as asymmetric mode switching~\cite{Doppler2016}. These achievements in photonics
suggest new possibilities for engineering novel
magnonic devices by taking advantage of non-Hermitian effects
already present in magnonic systems.

In this perspective we outline recent advances toward using non-Hermitian Hamiltonians to describe classical and quantum magnetic systems, a field that has grown rapidly in popularity in recent years and provides new insights into purely magnonic systems as well as hybrid magnon systems. In Sec.~\ref{Sec:Theory} we briefly outline the properties of non-Hermitian Hamiltonians that are most relevant for the study of magnetic systems. In Sec.~\ref{Sec:PureMagnon} we discuss manifestations of non-Hermitian physics in purely magnonic systems, ranging from dynamical and topological phase transitions associated with magnonic EPs to the emergence of the magnetic skin effect. In Sec.~\ref{Sec:Hybrid}, we present the current understanding and experimental results on exploring non-Hermitian physics in hybrid magonic systems, where magnons interact with other systems such as microwave photons. Finally, Sec.~\ref{Sec:Outlook} presents an outlook on the field of non-Hermitian magnonics and provides some directions for future research.

\section{Theory of Non-Hermitian Hamiltonians \label{Sec:Theory}}

The theory of non-Hermitian Hamiltonians and their properties is expansive, with more new results being published each year. Here, we review a few of the key properties of non-Hermitian Hamiltonians that are particularly relevant to magnonic systems. However, we do not provide an exhaustive review of non-Hermitian Hamiltonian frameworks. For more information on non-Hermitian formulations of quantum mechanics, see Refs.~[\onlinecite{Bender1998, Mostafazadeh2002, Mostafazadeh2002a, Mostafazadeh2002b}], for more information on non-Hermiticity and topology, we refer the reader  to Refs.~[\onlinecite{Kawabata2019},\onlinecite{Bergholtz2021}].
 
Consider a coupled two-mode system which can exchange energy with its environment. In many cases, one can define an effective non-Hermitian Hamiltonian for such a system, i.e. $\mathcal{H} = (\hat{a}^\dagger, \hat{b}^\dagger)H(\hat{a}, \hat{b})^T$ where 
\begin{equation}
    %\mathcal{H} = (\omega_a - i\gamma_a)a^\dagger a + (\omega_b -i\gamma_b)b^\dagger b + g(a^\dagger b + b^\dagger a)
    H = \begin{pmatrix}
    \omega_a - i\kappa_a & g \\ g & \omega_b - i\kappa_b
    \end{pmatrix}.
    \label{Eqn:H-nH-1}
\end{equation}
Here, $a^\dagger, a (b^\dagger, b)$ are, respectively, the creation and  annihilation operators for mode $a(b)$. Here and elsewhere in this manuscript we set $\hbar = 1$. In the context of magnonic systems, these operators are bosonic, i.e. they obey the bosonic commutation relations, although in general a similar fermionic model could also be studied. The coupling strength is parameterized by the complex number $g = \mathrm{Re}(g)+ i\mathrm{Im}(g)$,  $\omega_{a(b)}$ is the energy of the bare modes, and $\kappa_{a(b)} > 0 (< 0)$ indicates the damping (gain) rate for each mode. The modes can be two magnonic modes or two modes in a hybrid system, for example a magnon and a microwave photon.

The eigenvalues of Hamiltonian~\eqref{Eqn:H-nH-1} are in general complex, however additional properties of the eigenvalues and eigenvectors can be understood through the symmetry of $\mathcal{H}$. In the early 1990's it was realized that a class of Hamiltonians which obey $\PT$-symmetry can have real eigenvalues, and therefore a self-consistent quantum mechanical theory was developed which did not require the Hamiltonian to be Hermitian~\cite{Bender1998, Bender2007}. $\PT$-symmetry refers to a symmetry under parity and time-reversal operations, where $\mathcal{P}\mathcal{T}H = H\mathcal{P}\mathcal{T}$ for some Hamiltonian $H$. The parity operation is a unitary operator defined by its effect on position $\vec{r}$ and momentum $\vec{p}$ as $\mathcal{P}: \vec{r} \rightarrow -\vec{r} ; \vec{p}\rightarrow -\vec{p}$, while $\mathcal{T}$ is an anti-unitary operator defined by $\mathcal{T}: i\rightarrow  -i, \vec{r} \rightarrow \vec{r} ; \vec{p}\rightarrow -\vec{p}$. Typically $\mathcal{P} = \sigma_{x,y,z}$ where $\sigma$ are the Pauli matrices and $\mathcal{T}$ is defined by complex conjugation. 

While the intense focus on $\PT$-symmetric Hamiltonians was initially due to their ability to have a real spectrum,  $\PT$-symmetry only dictates that eigenvalues come in complex-conjugate pairs $(E, E^*)$, not that they are real. Therefore, systems described by a $\PT$-symmetric Hamiltonians can be in the `$\PT$-unbroken' regime, where the spectrum is real, or the '$\PT$-broken' regime, where complex eigenvalues emerge. A word on terminology here: `$\PT$-unbroken' refers to a situation in which \emph{both} the Hamiltonian and its eigenvectors $|E\rangle$ obey $\PT$-symmetry, whereas `$\PT$-broken' indicates that only the Hamiltonian obeys $\PT$-symmetry, while its eigenvectors do not. The ability to tune a system from a $\PT$-unbroken to $\PT$-broken regime has physical consequences, as $\PT$-symmetry breaking indicates an instability of one or more eigenmodes.

$\PT$-symmetry is not the only relevant symmetry in the theory of non-Hermitian Hamiltonians. Systems obeying anti-$\PT$-symmetry, where $\mathcal{P}\mathcal{T}H = - H\mathcal{P}\mathcal{T}$, also have interesting properties. Anti-$\PT$-symmetric Hamiltonians can be engineered in systems with loss and/or dissipative coupling. In contrast to the $\PT$-symmetric case, the spectrum is always complex and becomes completely imaginary when crossing an exceptional point in parameter space. Anti-$\PT$ symmetric systems can exhibit unconventional properties such as level-attraction~\cite{Zhao2020}, and may be useful for some sensing applications~\cite{Nair2021}.

Additionally, pseudo-Hermiticity is another important symmetry, defined by $\eta H^\dagger \eta^{-1} = H$ where $\eta$ is a unitary Hermitian operator with $\eta^2 = 1$. In a series of papers, A. Mostafazadeh rigorously showed that pseudo-Hermiticity, not $\PT$-symmetry, is the necessary symmetry condition to guarantee a real spectrum~\cite{Mostafazadeh2002, Mostafazadeh2002a, Mostafazadeh2002b}. Since these early works, others have shown that pseudo-Hermiticity and conjugated pseduo-Hermiticity (i.e. $\eta H^\dagger \eta^{-1} = H^*$) are crucial symmetries for understanding topological properties of non-Hermitian models~\cite{Kawabata2019, Bergholtz2021}. For example, Ref.~[\onlinecite{Lieu2018}] showed that breaking conjugated pseudo-Hermiticity leads to a topological phase transition in a 1D non-Hermitian lattice model while $\PT$-symmetry breaking does not dictate a topological transition. Furthermore, pseudo-Hermitian Hamiltonians naturally emerge in the theory of anti-ferromagnetic magnonics, for example as discussed in Refs.~[\onlinecite{Proskurin2017}, \onlinecite{Daniels2018}].

In addition to the expanded symmetry properties of non-Hermitian Hamiltonians, they also exhibit unique features in their energy spectra at degeneracy points. NH systems have two different types of degeneracies, (i) so-called `diabolic' points which have degenerate eigenvalues but orthogonal eigenvectors - these are the same as in Hermitian systems - and (ii) `exceptional points'  where both the eigenvalues and eigenvectors of $\mathcal{H}$ become degenerate. Near the exceptional point (EP), eigenvectors coalesce into a single point on the Reimann surface~\cite{Heiss2012}. This mathematical peculiarity of non-Hermitian systems also has physical consequences, namely the extreme sensitivity of the energy spectrum to perturbations, where for a small deviation $\epsilon \ll 1$ around an exceptional point, the energy levels have a splitting $\propto  \sqrt{\epsilon}$, in contrast to the linear splitting $\propto \epsilon$ seen at a diabolic point. Numerous efforts are underway to harness this property of non-Hermitian EPs for enhanced sensing~\cite{Chen2017}. Exceptional points in magnetic systems have been intensely investigated both theoretically~\cite{EPmagnon, Galda2016, Xu2017, Galda2019, Grigoryan2022} and experimentally~\cite{EP, Zhang2019b, Zhao2020, Zhang2021}.

To illustrate the emergence of an EP, we can consider a simpler version of Hamiltonian~\eqref{Eqn:H-nH-1} with equal energies $\omega_a = \omega_b = \omega$ and non-dissipative coupling i.e. $\mathrm{Im}(g) = 0$. In that case, we have 

    \begin{align}
    H=\begin{pmatrix} \omega & 0 \\ 0 & \omega\end{pmatrix} + \begin{pmatrix} - i \kappa_{a} & g \\ g & - i \kappa_{b} \end{pmatrix}\,
    \label{2}
    \end{align}
    
with eigenvalues
\begin{equation}
    \varepsilon_{\pm} = \omega -i\frac{\kappa_1 + \kappa_2}{2} \pm \frac{\sqrt{4g^2 - (\kappa_1 - \kappa_2)^2}}{2}
\end{equation}
    
The un-normalized eigenvectors are 
\begin{equation}
    \phi_\pm = \left(
    \frac{i(\kappa_2-\kappa_1) \pm \sqrt{4g^2-(\kappa_1 - \kappa_2)^2}}{2g},  1
    \right)^T
\end{equation}
where we can see that the EP emerges for $2g=\kappa_{1}-\kappa_{2}$.\\

An interesting and emerging field of research in non-Hermitian magnonic systems concerns higher-order EPs, which can be exceptionally sensitive to perturbations. In general, $N$ eigenvectors can coalesce around the same EP, which is therefore dubbed an $N$-th order EP. For a small deviation $\epsilon\ll1$ around an $N$-th order EP, the eigenvalues can be expanded in terms of a Puiseux series of order $\epsilon^{1/N}\gg\epsilon$. The $N$-th root singularity signals a drastic response of the system to a  perturbation around the EP, which has led to additional efforts in the development of EP-based sensors~\cite{Chen2017, Nair2021}. Furthermore, in systems with many tunable parameters there is the intriguing possibility of exploring exceptional lines or exceptional surfaces, where there is an extended parameter regime over which eigenvectors coalesce, providing additional flexibility in experimental set-ups~\cite{Grigoryan2022}.

\begin{figure*}[htbp]
    \centering
    \includegraphics[width=0.90\linewidth]{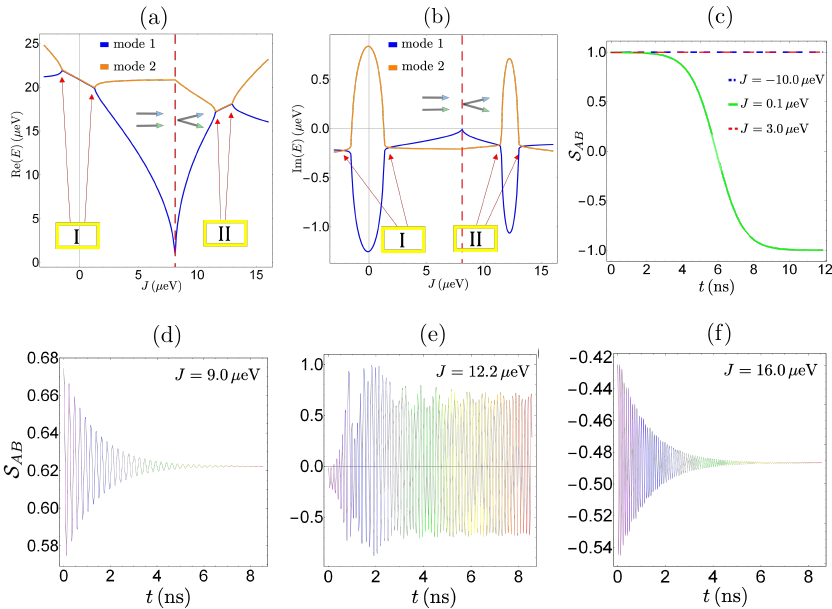}
    \caption{Real (a) and imaginary (b) energy spectra of the long-wavelength magnetization dynamics of a magnetic bilayer described by Eqs.~(\ref{LLG1}), (\ref{LLG2}) and~(\ref{effField}). The red dashed line separates a collinear from a non-collinear ground state. There are two regions enclosed by exceptional points, i.e., region I and II, which appear in correspondence of, respectively,  weak and strong antiferromagnetic interlayer coupling $J$. (c-f): The time evolution of $S_{AB}(t)=\mathbf{m}_{A}(t) \cdot \mathbf{m}_{B}(t)$  for different values of the interlayer coupling $J$. The FM-to-AFM dynamical phase transition emerges  in region I for small interlayer coupling, e.g., $J=0.1$ meV. Instead, for values of $J$ further away from region I, the relative alignment of the macrospins remains the one of the corresponding ground state~(c). A periodic dynamical phase emerges only within region II~(e). In each figure, the parameters are set to $B=0.1$ T, $K=45.9$ $\mu$eV, $\alpha_1=0.06$ and $\alpha_2=0.04$.  Figures (a-f) are adapted from Ref.~[\onlinecite{DengEP}].}
    \label{Fig2}
\end{figure*}

\section{Purely Magnonic Systems \label{Sec:PureMagnon}}

The key ingredient to engineering the non-Hermitian phenomena observed in photonic systems and metamaterials is the tunability of the parameters controlling the non-Hermiticity of an open system, i.e., gain and loss. 
Since the birth of spintronics,  magnetization dynamics have been known to be inherently lossy due to the ubiquitous spin non-conserving interactions with the crystalline lattice and other degrees of freedom. 
In the last decades, experimental techniques based on the injection of spin angular momentum via,  e.g.,  spin–transfer torques~\cite{SLONCZEWSKI1996L1,PhysRevB.54.9353,ralph2008spin,PhysRevB.54.9353,spinpumping,spinpumping1}, parametric pumping~\cite{bracher2017parallel,demokritov2006bose,sandweg2011spin}, thermal gradients~\cite{PhysRevB.81.214418,caloritronics,PhysRevB.93.100402} or coherent drives~\cite{Du2020,PhysRevB.94.214428,PhysRevB.100.064410}, have been established as controlled ways to induce effective gain of the magnetization dynamics. The feasibility with which the balance between gain and loss can be tuned seemingly makes magnetic systems promising solid-state platforms for realizing and controlling non-Hermitian phenomena. Besides, investigating magnetic systems through the lenses of the novel non-Hermitian theories might shed new
light on magnetic phenomena, e.g., magnetic topological phases, that have been explored only within Hermitian frameworks despite the lossy nature of the magnetization dynamics. 

While a comprehensive microscopic theory of dissipative magnetization dynamics is still lacking,  in the long-wavelength limit losses are conventionally accounted for via a phenomenological Gilbert damping parameter $\alpha$~\cite{Gilbert,Hickey2009}. It is within this framework that the connection between magnetic systems and non-Hermitian phenomena was first pointed out. In Ref.~[\onlinecite{EPmagnon}], the authors show that a system comprised of two exchange-coupled ferromagnetic films, whose long-wavelength dynamics can be described by a set of coupled Landau–Lifshitz–Gilbert (LLG) equations~\cite{Landau1}, i.e.,
\begin{align}
\frac{d \textbf{m}_{1}}{d t}= - \gamma \mathbf{m}_{1} \times \mathbf{H}_{1} - \gamma J \mathbf{m_{1}}\times \mathbf{m}_{2}+ \alpha_{1}\mathbf{m}_{1} \times \frac{d \textbf{m}_{1}}{d t}\,, \label{LLG1} \\
\frac{d\textbf{m}_{2}}{d t}= - \gamma \mathbf{m}_{2} \times \mathbf{H}_{2} - \gamma J \mathbf{m_{2}}\times \mathbf{m}_{1}- \alpha_{2}\mathbf{m}_{2} \times \frac{d\textbf{m}_{2}}{d t}\,, \label{LLG2}
\end{align}
obeys $\mathcal{PT}$-symmetry for $\alpha_{1}=\alpha_{2} \equiv\alpha$.  Here, $\mathbf{m}_{1 (2)}$ is the orientation of the magnetization of the first (second) layer, $\mathbf{H}_{1 (2)}$  the local effective magnetic fields, and $J$ the interlayer exchange coupling constant. It is essential to note that the Gilbert damping parameter $\alpha$ appears with the opposite sign in Eq.~(\ref{LLG2}), reflecting  gain of the magnetization dynamics of the second layer.  Introducing the parity operation ($\mathcal{P}$) as the interchange $\mathbf{m}_{1} \leftrightarrow \mathbf{m}_{2}$, which implies $\mathbf{H}_{1} \leftrightarrow \mathbf{H}_{2}$, one can easily see that Eqs.~(\ref{LLG1}) and~(\ref{LLG2}) are invariant under  a combined parity $\mathcal{P}$ and time-reversal $\mathcal{T}$ ($t \leftrightarrow - t$) operation. The $\mathcal{PT}$-symmetry is also evident  in the Hamiltonian formulation  of Eqs.~(\ref{LLG1}) and~(\ref{LLG2}), which can be obtained by linearizing the magnetization dynamics around the equilibrium orientation of the magnetic order parameters. Namely, by assuming $\mathbf{m}_{i}\simeq(\delta m_{i,x},\delta m_{i,y},1)$, with $|\mathbf{m}_{i}| \simeq 1$, Eqs.~(\ref{LLG1}) and~(\ref{LLG2}) can be recasted,  upon Fourier transformation, as a Schrodinger equation, i.e.,
\begin{eqnarray}
\mathcal{H}(\mathbf{k})\psi(\mathbf{k})=\epsilon(\mathbf{k})\psi(\mathbf{k}),
\end{eqnarray}
where $\epsilon(\mathbf{k})$ is the eigenergy of the eigenvector $\psi(\mathbf{k})=\left[\psi_1(\mathbf{k}),\psi_2(\mathbf{k})\right]^T$, with

\begin{eqnarray}
\psi_{i}(\mathbf{k})&=&\delta m_{i,x}(\mathbf{k})+i \delta m_{i,y}(\mathbf{k})\,.
\end{eqnarray}

The effective non-Hermitian (quadratic) Hamiltonian $\mathcal{H}(\mathbf{k})$ maps directly into Eq.~(\ref{2}): by varying the strength of non-Hermiticity, i.e., $\alpha$ (with $\alpha \ll 1$), or the interlayer coupling $J$, one can access a second-order exceptional point, which signals a transition  from a purely real to spectrum to one with complex conjugate pairs of eigenvalues. Yu \textit{et al.} generalized this approach to higher-order exceptional points by considering ferromagnetic trilayers consisting of a gain, a neutral, and a (balanced-)loss layer~\cite{Yu2020}. The linearized magnetic dynamics of this structure exhibit a third-order exceptional point, which, the authors argue,  might display a magnetic sensitivity three orders of magnitude higher than conventional magnetic sensors based on magnetic tunnel junctions.

Inspired by these theoretical proposals,  H. Liu and coauthors fabricated  a series of \textit{passive} magnonic  $\mathcal{PT}$-symmetric devices in the form of a trilayer structure with two magnetic layers, separated by a platinum (Pt) interlayer, whose  thickness $d$ was varied from one device to another~\cite{EP}. The linearized magnetization dynamics of the devices can be understood in terms of  Eq.~(\ref{2}):  the magnetic layers are governed by different damping parameters $\kappa_{1}$ and $\kappa_{2}$,  while the thickness of platinum modulates an  RKKY exchange interaction of strength~$g$ between them. The authors probed the EPs by measuring the eigenfrequencies and the damping rates of the collective dynamics of the trilayers, to find  a device where the coupling strength was equal to the critical value $2g=\kappa_{1}-\kappa_{2}$. 

As noted  by Galda and coauthors~\cite{Galda2016}, in the  $\mathcal{PT}$-unbroken phase a magnetic system exhibits physical properties seemingly equivalent  to those of Hermitian systems, such as a real energy spectrum and stationary states. Instead,  in the $\mathcal{PT}$-broken phase, the system displays a complex energy spectrum and nonconservative dynamics. Thus, the $\mathcal{PT}$-symmetry-breaking phase transition occuring at an exceptional point can be identified as a  dynamical phase transition, i.e., between  stationary and nonstationary dynamics. These considerations naturally bring up the question  of whether the presence of EPs, which surface from the linearized dynamics of Eqs.~(\ref{LLG1}) and~(\ref{LLG2}) for $\alpha \ll 1$,  has an impact as well on the non-linearized LLG dynamics and, if this is the case, whether exact $\mathcal{PT}$-symmetry might represent a strict requirement for a dynamical phase transition to  occur.

H. Yang and coauthors showed that Eqs.~(\ref{LLG1}) and~\eqref{LLG2}  display a ferromagnetic-to-antiferromagnetic (FM-to-AFM) dynamical phase transition for $\alpha_{1}=\alpha_{2} \equiv \alpha$ when the EP condition $J \sim \pm \alpha \omega_{0}$ is satisfied, where $\omega_{0}$ is the ferromagnetic  resonance frequency (FMR) of an individual macrospin~\cite{Yang2018}. Such transition can be easily understood by realizing that, since the Gilbert damping parameter is small, i.e., $\alpha \ll 1$, the EP condition translates into a very weak interlayer coupling $J$. 
 In this regime, the magnetic order parameters $\mathbf{m}_{1}$ and $\mathbf{m}_{2}$ are hardly coupled and each one eventually obeys its individual dynamics. Thus, the magnetic order parameter $\mathbf{m}_{1}$ relaxes to its static equilibrium orientation, while $\mathbf{m}_{2}$ switches, leading to an AFM orientation of the bilayer. 
 
Deng~\textit{et al.} showed that exact $\mathcal{PT}$-symmetry, i.e., $\alpha_{1}=\alpha_{2}$, is not required for the FM-to-AFM switching to occur: instead, while crossing an EP, a dynamical phase transition of the 
 coupled LLG dynamics can take place whenever one of the 
 magnetization dynamics experiences net gain~\cite{DengEP}, as shown in Fig.~\ref{Fig2}(c). 
In this work, the authors investigated also the role of easy-plane 
 anisotropies by setting the effective fields $\mathbf{H}_{1(2)}$ in Eqs.~(\ref{LLG1}) and~(\ref{LLG2}) to
 \begin{align}
\gamma \mathbf{H}_{1(2)} \rightarrow \gamma \mathbf{B}_{0} + 2 (K/\hbar) \mathbf{m}_{z, 1(2)}\,,
\label{effField}
 \end{align}
 where $\mathbf{B}=B_{0} \hat{\mathbf{x}}$ is a static magnetic field  and $K$ parameterizes the strength of an easy-plane anisotropy.  The results reveal  that in the  AFM region of interlayer coupling, i.e., $J>0$, a further region enclosed by two EPs appears, as shown in Figs.~\ref{Fig2}(a) and (b).
As displayed in Figs.~\ref{Fig2}(d-f), crossing one of these EPs also yields to  dynamical phase transition of the coupled LLG dynamics, i.e., from a damped to large-amplitude oscillatory regime, which can be understood in terms of a supercritical Hopf bifurcation.  

The interplay between EPs and magnetic dynamical phase transitions has yet to be experimentally explored. Promising platforms are offered by  synthetic antiferromagnets and van der Waals 2$d$ layered magnets~\cite{natureAFM,vdWreview}. Many of these systems support acoustic and optical magnon modes whose interactions could be  continuously tuned via a symmetry-breaking magnetic field or dynamic dipolar interactions, providing a further experimental knob  to probe these dynamical phase transitions~\cite{Sud2020,Shiota2020,PhysRevLett.123.047204,PhysRevApplied.15.044008,jeffrey2021effect}. 

So far, this section has focused on \textit{local} magnetic dissipation and gain as sources of non-Hermiticity. However, non-local dissipation and gain arising from spin pumping are known to play an important role in the magnetization dynamics of heterostructures, in particular at metallic interfaces. The interfacial damping-like spin pumping torque contributes to the LLG magnetization dynamics as~\cite{spinpumping,spinpumping1,spinpumping2}
\begin{align}
 \frac{d\mathbf{m}_{i}}{dt}= ... + \alpha_{d}\; \mathbf{m}_{i} \times \left( \mathbf{m}_{i} \times \frac{d\mathbf{m}_{i}}{dt} - \mathbf{m}_{j} \times \frac{d\mathbf{m}_{j}}{dt} \right) \times \mathbf{m}_{i}\,,
 \label{mixcond}
\end{align} 
where $i,j$ label adjacent magnetic layers and $\alpha_{d}$ is a parameter proportional to the real part of the spin-mixing conductance. Reference~[\onlinecite{Tser2020}] investigates the dynamics of a one-dimensional (1$d$) array of macrospins exchange-coupled through an isotropic medium. This model can be seen as an extension of Eqs.~(\ref{LLG1}) and~(\ref{LLG2}) to a multilayer structure in which each layer is lossy, and the damping-like spin pumping torques~\eqref{mixcond}, mediated by  spacer layers, are accounted for.  Reference~[\onlinecite{Tser2020}] shows that tuning both local or non-local damping can lead to level attraction between magnon modes, and thus EPs,  at finite momenta, both in ferromagnetic and antiferromagnetic spin chains. These results suggest that non-local dynamical interactions between magnets can provide a further knob for non-Hermitian band engineering.

In the vicinity of an EP, magnon spectral features can be drastically modified \cite{Tser2020}, which, in turn, can affect the the properties  routinely probed in spintronics setups, such as, e.g., spin conductivity and spin diffusion length. However, the systems discussed  so far display EPs   only at isolated momenta, i.e., finite momenta in the Brillouin zone (BZ) or at the BZ's center, i.e.,  $|\mathbf{k}|=0$, corresponding to the long-wavelength limit of magnetization dynamics. It is not likely that the presence of a single EP will influence  system’s properties, such as, e.g., transport coefficients, that depend on integrals over a large number of momenta. A recent work showed that a multitude of EPs in the Brillouin zone can emerge in systems with higher dimensionality, e.g., a van der Waals bilayer, when the balance between gain and loss of the layers is conveniently tuned~\cite{Li2022}. It is not yet clear how the presence of EPs can affect, e.g., spin transport. To gain insights on this issue, future studies should address the inclusion of non-Hermitian spectral singularities in the Green's function formalism of spin transport.

The non-Hermitian nature of the LLG magnetization dynamics has a profound effect not only on its spectral singularities, but also on its topological phases. Magnetic systems have been extensively explored as potential platforms for realizing bosonic analogs of the topological band structures that have become ubiquitous in electronic systems~\cite{McClarty}. Most of the theoretical proposals, however, have relied on Hermitian models, due to the lack of a framework for addressing dissipative topological phases. While non-Hermitian topological theories have emerged only relatively recently, this field has gained very strong momentum over the past few years~\cite{Kawabata2019,Gong2018,ashida2020non,Bender2007} and has the potential to shed light on the topological properties of open  systems, including magnonic ones.
 
To avoid ambiguities, it is important to remark that non-Hermitian Hamiltonians have emerged in some early works on topological magnonic phases\cite{Shindou2013, Shindou2014}, as well as in the theory of antiferromagnetic magnonics~\cite{Proskurin2017, Daniels2018}. This is because a generic $N \times N$ bosonic Bogoliubov-de-Gennes (BdG) Hamiltonian $\mathcal{H}_{\text{BdG}}$ has to be diagonalized by a paraunitary matrix that preserves the canonical bosonic commutation relations, which amounts to the diagonalization of the non-Hermitian matrix $\mathcal{H}_{\text{nH}}=\sigma_{3} \mathcal{H}_{\text{BdG}}$, with 
\begin{align}
\sigma_{3}=\begin{pmatrix} 1_{N\times N} & 0 \\ 0 & -1_{N\times N} \end{pmatrix}\,.
\end{align}
To identify  the topological phase of the Hamiltonian $\mathcal{H}_{\text{nH}}$,  one needs to invoke the theory of non-Hermitian topological classifications, while  neglecting particle-hole symmetry~\cite{Kawabata2019,kondo}. The latter  is meaningless in a system of free bosons obeying Bose-Einstein statistics, i.e., no state is fully occupied and, thus,  the concept of filling magnon energy bands up to a given energy level does not make sense. Importantly, if the system of  free bosons is energy-conserving, i.e., $\mathcal{H}_{\text{BdG}}=\mathcal{H}^{\dagger}_{\text{BdG}}$, the Hamiltonian $\mathcal{H}_{\text{nH}}$ is  pseudo-Hermitian and displays a real energy spectrum.  

In what follows, we will not discuss such BdG Hamiltonians. Instead, we will focus on the topology of open dissipative or dissipative-driven magnonic systems, in which the non-Hermiticity stems from coupling to the environment. 
 While the  field of non-Hermitian topological theories is still actively growing, a model that has been thoroughly understood is the $\mathcal{PT}$-symmetric non-Hermitian Su–Schrieffer–Heeger~\cite{Lieu2018}, described by the Hamilltonian
\begin{align}
\mathcal{H}_k=\begin{pmatrix} i u &  J + \tilde{J} e^{-ik} \\ J + \tilde{J} e^{ik} & - iu \end{pmatrix}\,.
\label{169}
\end{align}
Here, $J$ and $\tilde{J}$ are the staggered  hopping amplitudes of the one-dimensional ($1d$) tight-binding chain and $u$ parameterizes an imaginary staggered potential. 
The model~(\ref{169}) has $\mathcal{PT}$-symmetry and thus two parameter regimes, i.e., the $\mathcal{PT}$-broken and unbroken phases, separated by an EP. The unbroken phase is defined by a fully real spectrum under periodic boundary conditions, while in the broken phase eigenvalues of Eq.~(\ref{169}) come in complex conjugate pairs. Similarly to its Hermitian counterpart, Eq.~(\ref{169}) undergoes a topological transition characterized by the emergence of (a pair of) topologically nontrivial edge states.

Interestingly, in the $\mathcal{PT}$-unbroken phase,  $\mathcal{PT}$-symmetry is spontaneously broken by the boundaries. Consequently, the energies of the edge states come as complex-conjugate pairs, i.e., there is a  ``lossy" and a ``lasing" edge state, whereas the bulk states remain stable, as the bulk displays a purely real spectrum. Such edge states have been probed in  several experiments in both $\mathcal{PT}$-active and passive photonic systems~\cite{lasing1,lasing2,lasing3}.
The topologically nontrivial phase is characterized by a topological invariant, the global Berry phase~\cite{liang2013topological}, which can be nonzero  both in the $\mathcal{PT}$-broken and $\mathcal{PT}$-unbroken region. While both can be topologically nontrivial, these phases are profoundly different: in the former, the edge states' energies come as a complex-conjugate pair, while the bulk states are purely real; in the later, the bulk bands are merged into complex conjugate pairs and the edge states are no longer separated from the bulk states. As we will discuss later in detail, such distinction can profoundly affect the dynamics of the system and the experimental detection of the topological edge states.

Inspired by the model~(\ref{169}), Flebus \textit{et al.} proposed the first magnonic realization of a non-Hermitian topological phase~\cite{Flebus2020}. The authors considered an array of spin-torque oscillators (STOs), i.e., current-driven magnetic nano-pillars coupled by metallic spacers. The latter mediate a (weak) staggered RKKY interaction, i.e., $J$ and $\tilde{J}$ in Eq.~(\ref{169}), which can be realized by modulating the spacers' length or composition.
The dissipation of magnetization dynamics is an inherent property of the magnetic nanopillars and is captured via a Gilbert damping parameter within the LLG formalism. The gain at each odd-site STO is provided by an external bias, i.e., a spin-transfer torque exerted by a spin-polarized current impinging on the free ferromagnetic layer of the STO. Focusing on the linear regime of fluctuations and for a spin current $J_{s}=2\alpha \omega_{0}$, where $\omega_{0}$ the FMR frequency  of an isolated STO, the system can be mapped into Eq.~(\ref{169}) by identifying $u \equiv\alpha \omega_{0}$. 

In order to shed light on the interplay between  non-Hermitian topology and non-linear spin dynamics, the authors performed numerical simulations of the LLG dynamics of an array of twenty STOs. Their results show that, in the topologically nontrivial phase with unbroken $\mathcal{PT}$-symmetry, the STO located at the left edge, which hosts the ``lasing" magnon edge state, starts precessing, while the bulk STOs remain inactive. Over long times, the system emits only a topologically-protected microwave signal at the left  edge, yielding a clear experimental signature of the topological phase of the $\mathcal{PT}$-symmetric Hamiltonian~(\ref{169}). In the topologically trivial phase, instead, all driven STOs start precessing simultaneously, as one would expect from a classical standpoint. In the $\mathcal{PT}$-broken topologically nontrivial phase, the STO at the left edge starts precessing, but, soon after, the other STOs driven by spin-current injection start precessing as well.

The authors explored as well the effects of perturbations relevant to experimental implementations of the system~\cite{Flebus2020,Gunnink2022}. The metallic spacers might mediate spin pumping~(\ref{mixcond}) or the injected spin current might not yield exact balance of gain/loss, i.e., $J_{s}\neq 2\alpha \omega$ (with $|J-\tilde{J}|>J_{s}/2$). In both cases, the authors find that the edge states are still protected by a symmetry, i.e., chiral-inversion and chiral, respectively. In the absence of spin pumping and for $J_{s}<2\alpha \omega$, the dynamics of the system maps into the  $\mathcal{PT}$-unbroken topological phase.  For larger spin currents or finite spin pumping,  the edge STO starts precessing before the bulk ones, analogously to the $\mathcal{PT}$-broken phase. A similar dynamics is realized in the presence of dipolar interactions and for finite temperature, as long as the perturbations are weak enough.

Extending this approach to 2-dimensional (2$d$) models, Li and coauthors find that, in magnetic bilayers, it is not possible to realize a $\mathcal{PT}$-symmetric phase away from the long-wavelength limit where the nontrivial topology stems from a chiral spin interaction, e.g., the Dzyaloshinskii–Moriya interaction (DMI)~\cite{Li2022}.  For an exact balance of gain and loss between the bottom and top layer, the system becomes $\mathcal{PT}$-symmetric only in the absence of DMI, and thus of topological edge states. In the presence of DMI, the systems acquires a peculiar wavevector-dependent pseudo-Hermiticity, which is not broken by the boundaries. Thus, for a small degree of non-Hermiticity (i.e., $\alpha \ll1$), both the bulk and the edge state spectra can be found to be purely real. Similarly, for a higher strength of non-Hermiticity, the eigenenergies of both bulk and edge states come as complex-conjugate pairs. While non-Hermitian topological insulators with entirely real spectra have long been sought after~\cite{Kawabata2020,Hu2011}, the authors argue that this regime might not be ideal for identifying topological signatures of bosonic systems. Because of Bose statistics, it is not possible to thermally occupy the edge states without populating lower-energy bulk states as well. The strategy offered by Ref.~[\onlinecite{Flebus2020}] appears more promising, i.e., if the lower-energy bulk states are purely real while an edge state has finite positive imaginary lifetime, the topological edge state dynamics will anticipate the bulk ones.  However,  topological non-Hermitian magnonic systems in 2$d$ or higher dimensions displaying both lasing edge states and a purely real bulk spectrum  have yet to be found.

\begin{figure}
\centering
{{\includegraphics[trim=0cm 0cm 0cm 0cm, clip=true,width=8.5cm, angle=0]{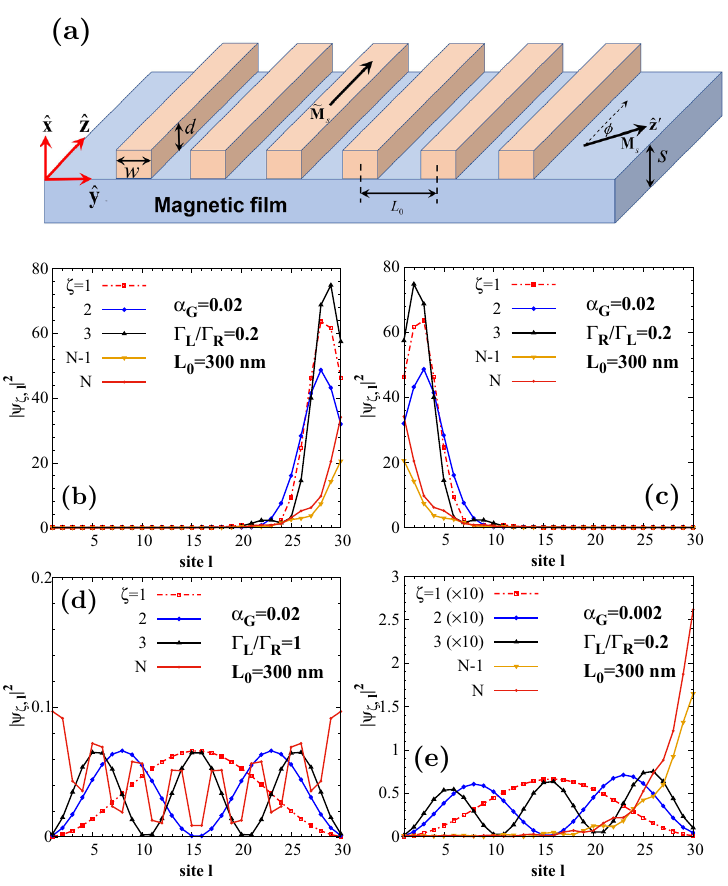}}}
\caption{(a)  Set up of Ref.~[\onlinecite{PhysRevB.105.L180401}]: a periodic array of magnetic nanowires on top of a thin magnetic film. The direction of the saturated magnetization $\tilde{\mathbf{M}}_{s}$ of the wire is pinned along the $\hat{\mathbf{z}}$ direction, while the saturated magnetization $\mathbf{M}_{s}$ of the film is tunable by the applied magnetic field in the film plane. By changing the relative orientation of the magnetizations, it is possible to tune the  relative strength with which the wire Kittel mode couples with, respectively, the right- and left-going spin waves in the thin film. This results in a chiral effective coupling between the magnetic nanowires, parametrized by the coupling strength $\Gamma_{L}$ ($\Gamma_{R}$) from the left to right (right to left) wires.  (b-e): Distribution of the probability density  of the normalized eigenmodes $\psi_{\zeta,l}$ under different conditions, with $\zeta$ ($l)$ labeling the mode (site).  All the modes are localized at the edge in (b) and (c) when the coupling is chiral, i.e., $\Gamma_{L}/\Gamma_{R} \neq 1$, and the Gilbert damping $\alpha_{G}$ of the magnetic film is relatively strong.  When $\Gamma_{R}>\Gamma_{L}$  all the modes are localized at the right
edge~(b), but become localized at the left edge
when the chirality is reversed with $\Gamma_{L}>\Gamma_{R}$. The skin modes vanish either without chirality (d) or with weak film damping (e). (e) The probability density of the modes $\zeta=1, ..,3$ is enhanced by an order of magnitude ($\times 10$) in the plot. Figures (a-e) are adapted from Ref.~[\onlinecite{PhysRevB.105.L180401}].}\label{Fig3}
\end{figure}

Up to now, our discussion has focused on examples of non-Hermiticity captured by  the  LLG formalism, i.e., phenomenologically and in the long-wavelength limit.  The proper inclusion of non-Hermiticity, i.e., dissipation and/or gain, far away from the long-wavelength limit is a far less explored problem. 
The physics of dissipative interactions in  magnetic systems is very complex and the effective magnon lifetime stems from a variety of spin non-conserving mechanisms, e.g., interactions of magnons with phononic and electronic degrees of freedom, and magnon scattering on extrinsic impurities. As pointed out by McCartly \textit{et al.}~\cite{McClarty2019}, it is not even necessary to invoke interactions with external degrees of freedom to obtain an effective non-Hermitian description of the magnetization dynamics.  When magnon-magnon interactions are considered,  the decay of one-magnon states into a multi-magnon continuum supplies a natural separation into system and bath.  

Spin-non-conserving interactions due to interactions with the crystalline lattice are obiquitous and represent a main source of magnetization dissipation. Several theoretical works have  addressed this problem by  providing approximate expressions for the magnon relaxation time. These expressions are, however, often given in the continuum limit and can not be readily incorporated in a lattice model with translational symmetry.
Reference~[\onlinecite{SkinMag}] partially addresses  this problem by developing generic phenomenological approach to describing magnetic dissipation within a lattice model. Namely, the authors propose to complement a Hermitian magnon model with a non-Hermitian component that can be extrapolated from \textit{ab initio} or experimental data. As a case study, Deng \textit{et al.} focus on the tight-binding model of a honeycomb ferromagnetic lattice with  spin-orbit interactions and incorporate the dissipation due magnon-phonon interactions according to the \textit{ab initio} results of Ref.~[\onlinecite{wang2021magnon}]. Their results show that a non-Hermitian magnetic system can display the skin effect. 
This effect, unique to non-Hermitian systems, amounts to  the localization of a macroscopic fraction of bulk eigenstates at the boundaries of non-Hermitian systems in which conventional  bulk-edge correspondence does not hold~\cite{PhysRevLett.121.086803}. Experimentally, it has been uncovered in $1d$ photonic systems and metamaterials with judiciously engineered non-Hermitian interactions~\cite{helbig2019observation,Weidmann2020,ghatak2019observation}, but not yet in magnetic systems. 

Mathematically, the emergence of skin effect in $2d$ can be understood via the spectral area law proposed by Zhang and coauthors~\cite{Zhang2022}. When the imaginary part of the spectrum can not be written as function of the real part (or \text{vice versa}), the periodic-boundary-conditions spectrum covers a finite area on the complex plane. In two dimensions, this implies that the mapping between momenta and energy becomes $2d$ to $2d$, contrary to the conventional Hermitian $2d$ to $1d$ correspondence.  Within a $2d$ to $1d$ ($2d$) momentum-energy mapping, for a wave impinging at the boundary there are infinite (finite) reflection channels, and an open boundary eigenstate can (can not) be described as superposition of Bloch waves, as discussed in detail in Ref.~[\onlinecite{Zhang2022}].
However, from a physical point of view, the microscopic mechanisms and time-scales underlying the emergence of the magnetic skin effect are not yet fully understood. While the results of Ref.~[\onlinecite{SkinMag}]
suggest that the interplay between chiral spin interactions and nonlocal magnetic dissipation plays a key role,  a comprehensive study of the full quantum dynamics appears necessary to shed further light on this phenomenom.

Recently, in a theoretical work, Yu and coauthors~\cite{PhysRevB.105.L180401}  reported the emergence of the skin effect in a $1d$ array of magnetic nanowires deposited on a thin magnetic film, sketched in Fig.~\ref{Fig3}(a). In 1$d$, the non-Hermitian skin effect is far less subtle than its higher-dimensional generalizations as it arises from the asymmetry between left- and right-hopping probabilities, which naturally results in an accumulation of a macroscopic number of modes at one edge of the system. In Ref.~[\onlinecite{PhysRevB.105.L180401}], the source of chirality lies in the dipolar interactions between each nanowire and the spin waves traveling in the thin magnetic film. The interaction depends strongly on  on the relative orientation of the nanowire and thin film magnetizations: when the magnetizations of the wire and film are parallel (anti-parallel), the wire Kittel mode only couples with the right-going (left-going) spin waves~\cite{chiral1,chiral2,chiral2,chiral4}. Integrating out the film degree of freedom yields an effective non-Hermitian Hamiltonian for the wires,  whose exact diagonalization can reveal the accumulation of eigenmodes at one edge, as shown in Figs.~\ref{Fig3}(b-e). Figure~\ref{Fig3}(e) shows that the chirality itself is not sufficient to trigger a macroscopic accumulation of modes at one edge since the strong interference brought by the wave propagation is detrimental for accumulation. However, the latter can be suppressed by increasing the damping of the traveling spin waves: by choosing a magnetic film with larger damping,  all the modes are localized at one edge, as shown in Fig.~\ref{Fig3}(b-c). 

\section{Hybrid Magnonic Systems \label{Sec:Hybrid}}

Controllably coupling magnons to other degrees of freedom provides an expanded opportunity to explore non-Hermitian physics in hybrid quantum and classical systems. Magnons are a good candidate for integration into hybrid systems because they are ubiquitous in solid-state systems whenever spin is a relevant degree of freedom. Furthermore, both classical and quantum states of magnons can be generated and finely tuned using external sources such as magnetic field gradients and microwave photons. Several recent review articles highlight the importance of hybrid magnonic systems in both basic research and engineering applications~\cite{Lachance2019, Li2020, Awschalom2021, Yuan2022}. Here, we focus on hybrid systems designed specifically to study non-Hermitian phenomena and discuss current results and opportunities for future work. 

We can describe a hybrid quantum system containing $M$ magnon modes and $N$ external (i.e. non-magnonic) modes by the following non-Hermitian Hamiltonian (setting $\hbar = 1$)

\begin{align}
    \mathcal{H} = \sum_{\ell = 1}^N &(\omega_\ell -i\kappa_\ell)\hat{a}^\dagger_\ell \hat{a}_\ell + \sum_{j = 1}^M(\Omega_j -i\gamma_{j})\hat{m}^\dagger_j\hat{m}_j \nonumber \\
    &+ \sum_{\ell j}g_{\ell j}\hat{a}^\dagger_{\ell}\hat{m}_j + \tilde{g}_{j\ell} \hat{a}_\ell \hat{m}^\dagger_j.
    \label{Eqn:HybridHamiltonian}
\end{align}
The bosonic operators $\hat{m}^\dagger_j(\hat{m}_j)$ describe the creation (annihilation) of a magnon in mode $j$ and obey the commutation relations $[\hat{m}_j,\hat{m}^\dagger_{j'}] = \delta_{jj'}$. Note that $\hat{m}$ may describe magnons in the semiclassical Holstein-Primakoff approximation or quantum magnons, for example in low-temperature experiments. The operators $\hat{a}^\dagger_\ell(\hat{a}_\ell)$ describe the creation (annihilation) of an excitation in a non-magnon mode. In the systems we will consider here, the operators $\hat{a}$ also describe bosonic excitations with commutation relation $[\hat{a}_\ell,\hat{a}^\dagger_{\ell'}] = \delta_{\ell \ell'}$. Depending on the system under consideration, these excitations could be phonons, microwave photons, qubit levels, etc. The parameters $\omega_\ell$, $\Omega_j$ denote the resonance frequency of the non-magnon and magnon modes, respectively, and $\gamma_\ell$, $\kappa_j$ refer to the damping ($\gamma, \kappa > 0$) or gain ($\gamma,\kappa < 0$) rate of the magnon and non-magnon modes. The second line of Eq.~\eqref{Eqn:HybridHamiltonian} denotes the interaction between magnon and non-magnon excitations, with interaction strengths $g_{\ell j}$ and $\tilde{g}_{j \ell}$, where $g$ and $\tilde{g}$ may be complex valued. An interaction term of this form is typically valid within the rotating wave approximation which is valid in the near-resonant regime, $\omega \sim \Omega$.

\begin{figure*}[htbp]
    \centering
    \includegraphics[width=0.90\linewidth]{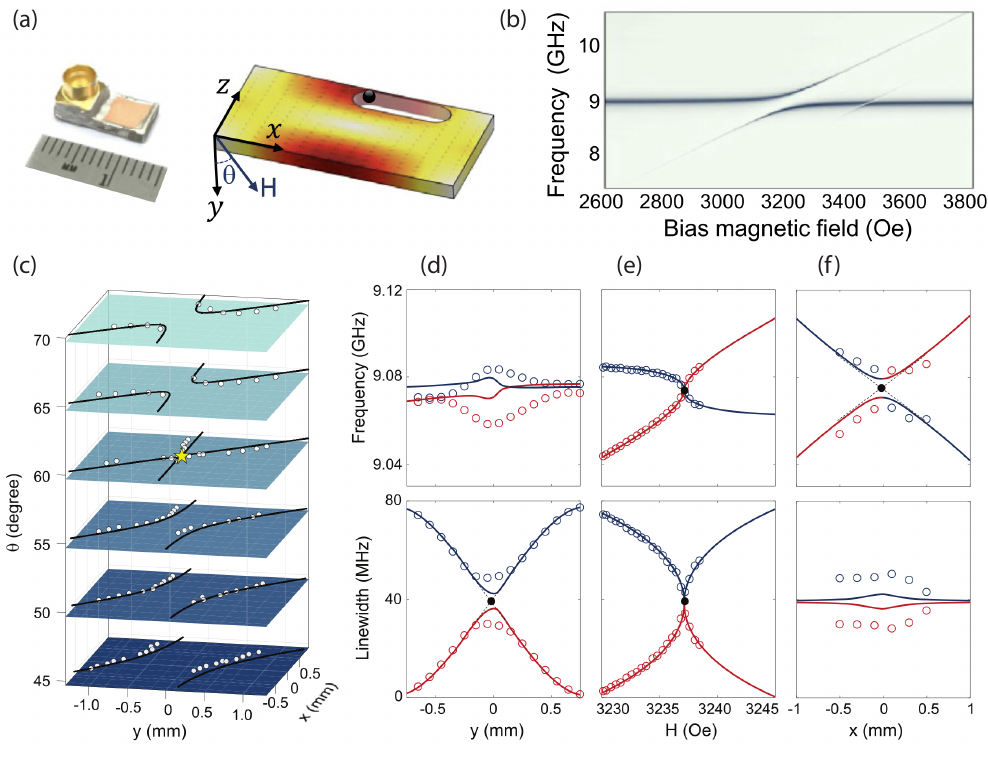}
    \caption{(a)~Typical cavity magnonics set-up, here shown from Ref.~[\onlinecite{Zhang2019b}]. A YIG sphere is place in a microwave cavity such that it interacts with the TE$_{101}$ magnetic field mode. The color scale shows the simulated magnetic field, and the slot for the YIG sphere is places near the field maximum. A bias field $H$ is placed in the $x-y$ plane with a tunable angle $\theta$. The actual microwave cavity is shown at left.~(b)~Cavity reflection spectra with changing bias field, showing a level-splitting around $3200$~Oe. The gap between levels indicates the strong coupling regime reachable in cavity magnonics.~(c)~ Cross-sections of the exceptional surface in a 3D $(x, y, \theta)$ parameter space. White circles are experimental results constructed from reflection spectra and black lines are numerical fitting. The yellow star indicates an exceptional saddle point.~(d)-(f) 1D cross-sections of the eigenfrequencies around the exceptional saddle point for varying parameters $x$, $H$, and $y$. Top row: Extracted resonance frequency (real part) and linewidth (imaginary part) show coalescence at the exceptional points (black dots) Circles are experimental data and solid lines are numerical calculations. Figures (a-f) are adapted from Ref.~[\onlinecite{Zhang2019b}].}
    \label{Fig:hybrid}
\end{figure*}

Hamiltonian~\eqref{Eqn:HybridHamiltonian} is manifestly non-Hermitian in several distinct ways. Clearly, the damping or gain described by $\gamma, \kappa$ breaks Hermiticity by introducing imaginary on-site terms. However, the interaction strengths $g$, $\tilde{g}$ can also break Hermiticity in situations where $\tilde{g}_{j \ell} \neq g_{\ell j}^*$. This latter feature is perhaps more straightforward to implement in hybrid magnonic systems, compared to other platforms, because the interactions can be dissipative. This makes hybrid magnonic systems excellent candidates for exploring a variety of non-Hermitian Hamiltonians.

Many studies of non-Hermitian physics in hybrid systems have been carried out using cavity magnonics, where a ferromagnetic material, typically a yttrium-iron-garnet (YIG) sphere, is placed in a microwave cavity, as shown in Fig.~\ref{Fig:hybrid}~(a). YIG is an ideal material for these experiments due to its low intrinsic damping and high spin density. The uniform magnon (Kittel) mode in the sphere hybridizes with microwave photons to form a magnon-polariton quasiparticle. Other realizations of cavity magnonics use nitrogen-vacancy centers in diamond~\cite{Zhang2021}. These systems are ideal for studying non-Hermitian Hamiltonians of the form of Eq.~\eqref{Eqn:HybridHamiltonian} because most of the parameters are highly tunable in experiments. 

The magnon frequency $\Omega$ is set by an external, static magnetic field. The cavity loss or gain rate $\kappa$ is set by precisely engineering the microwave input ports. The coupling strengths $g, \tilde{g}$ are tunable in a number of ways, for example by the angle of the external magnetic field~\cite{Zhang2019b} or the position of the YIG sphere in the cavity~\cite{Zhang2017}. These systems can enter the strong coupling regime where $g, \tilde{g} \gg \gamma, \kappa$~\cite{Zhang2014, Tabuchi2014, Zhang2019b}, as evidenced by the mode hybridization in cavity reflection spectra. An example of an experiment in the strong-coupling regime is shown in Fig.~\ref{Fig:hybrid}~(b). Furthermore, the interaction between magnons and photons can even be tuned between coherent ($g$ real) and completely dissipative ($g$ imaginary) regimes~\cite{Harder2018}, leading to a number of interesting non-Hermitian effects such as level attraction~\cite{Harder2018, Li2022b} and bistability~\cite{Yang2021}. The plethora of theoretical and experimental results in the last few years speak to the versatility of the cavity magnonics platform~\cite{Tabuchi2014, Zhang2014, Zhang2015, Zhang2017, Harder2017, Harder2018, Zhang2019, Zhang2019b, Zhao2020, Yu2020magnon, Yang2021, Zhang2021, Ren2022, Grigoryan2022, Li2022b}. For a thorough review of cavity magnonics, the reader is referred to Ref. [\onlinecite{Harder2021}] and references therein.

Most cavity magnonics setups operate in the two mode regime (i.e. $M + N = 2$), although larger systems have also been studied for example by Zhang et. al.~\cite{Zhang2015} who experimentally realized a system with $N=1$ and $M=8$. Multiple magnon and cavity modes can be introduced in a number of ways by adding additional YIG spheres to a single cavity~\cite{Zhang2015, Zhao2020}, or by coupling cavities containing individual YIG spheres into a larger array~\cite{Ren2022}.

Non-Hermitian Hamiltonians exhibiting many different types of symmetry have been experimentally realized in cavity magnonics. In Ref.~[\onlinecite{Zhang2017}] a single photon ($N = 1$), single magnon ($M = 1$) system was engineered such that $\Omega = \omega$ and $\gamma = -\kappa$ resulting in the $\PT$-symmetric Hamiltonian 
\begin{equation}
    \mathcal{H} = (\omega + i\gamma)\hat{a}^\dagger\hat{a} + (\omega -i\gamma)\hat{m}^\dagger\hat{m} + g(\hat{a}^\dagger\hat{m} + \hat{a}\hat{m}^\dagger).
\end{equation}
The EP at $g = \gamma$ was observed by tuning coupling $g$ across the transition from the $\PT$-unbroken to $\PT$-broken regime. In later works, similar $\PT$-symmetric setups have been employed with additional tunable parameters, resulting in higher dimensional exceptional lines~\cite{Grigoryan2022} and surfaces~\cite{Zhang2019}. Note that to reach the $\PT$-symmetric regime one of the modes must experience an effective gain. Here, the gain is achieved for the cavity mode using coherent perfect absorption~\cite{Zhang2017}. 

Pseudo-Hermiticity without $\PT$-symmetry may be realized in a three mode cavity magnonic system, created using either one cavity mode and two magnon modes ($N=1$, $M=2$)~\cite{Zhang2019} or vice versa, ($N=2$, $M=1$)~\cite{Harder2017, Zhang2020}. Multiple second-order EPs have been observed in such a system~\cite{Harder2017}, which could lead to expanded mode-switching pathways in the parameter space. Since they include more than two modes, these systems are also good candidates for observing higher-order EPs. Third order EPs have been predicted for the $M=1$, $N=2$ system, created using two YIG spheres in one cavity, but have yet to be observed experimentally~\cite{Zhang2019}.

Cavity magnonics experiments with additional tunable parameters have the ability to further explore exceptional surfaces and exceptional saddle points. Consider the Hamiltonian Eq.~\eqref{Eqn:H-nH-1}, which was experimentally realized in Ref.~[\onlinecite{Zhang2019}]. By tuning the Hamiltonian parameters individually, the exceptional point can be reached in different ways. In this experiment, the position of the YIG sphere was tunable in the $x-y$ plane. Additionally, a bias magnetic field applied in the $x-y$ plane had a tunable magnitude $H$ and angle $\theta$. This enabled realization of a 4D parameter space $(x, y, H, \theta)$ for the Hamiltonian, leading to experimental observation of an exceptional surface where individual exceptional points occur in many places in parameter space. As shown in Fig.~\ref{Fig:hybrid}~(c)-(f), the real and imaginary parts of the eigenfrequencies of Eq.~\eqref{Eqn:H-nH-1} are extracted from the cavity reflection spectra by finding the resonant frequency and linewidth, respectively. Near an exceptional point, the reflected modes coalesce. 

\begin{figure}[th]
    \centering
    \includegraphics[width=0.9\columnwidth]{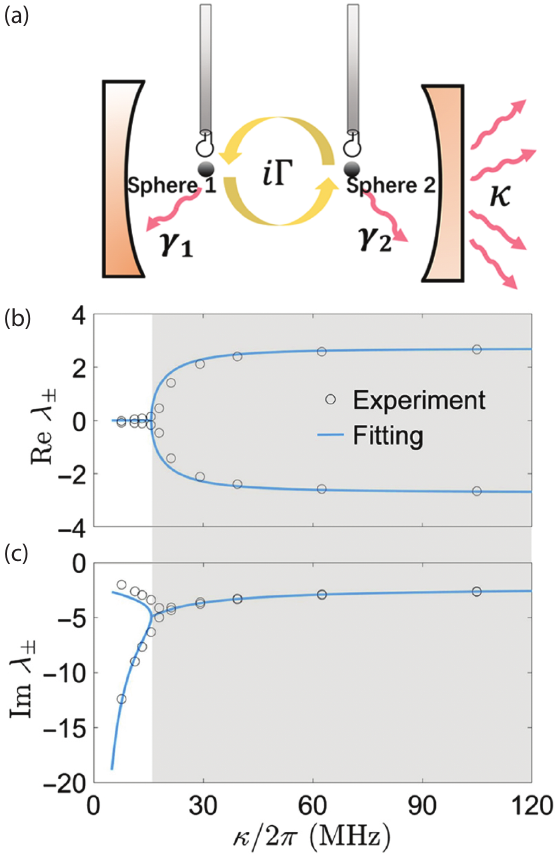}
    \caption{Cavity magnonics set up engineered to realize an Anti-$\PT$ symmetric Hamiltonian.~(a)~Schematic of the experimental system. A large cavity dissipation rate $\kappa$ yields an effective dissipative coupling $i\Gamma$ between 2 YIG spheres in a microwave cavity. $\gamma_1, \gamma_2$ are the individual damping rates of the YIG spheres.~(b)-(c)~Observation of the Anti-$\PT$ phase transition. The real (b) and imaginary (c) part of the eigenvalues of the anti-$\PT$ symmetric Hamiltonian. As the cavity dissipation $\kappa$ increases, the eigenvalues go from purely imaginary to having a non-zero real part, indicating the anti-$\PT$ transition. Figs. (a-c) are adapted from Ref.~[\onlinecite{Zhao2020}]}
    \label{Fig:Anti-PT}
\end{figure}

Anti-$\PT$-symmetry has also be experimentally realized in a system with $M = 1$, $N=2$ with a lossy cavity where $\kappa \gg \gamma$~~\cite{Zhao2020}. In this regime, the cavity mode can be adiabatically eliminated, resulting in a purely imaginary magnon-magnon coupling as shown schematically in Fig.~\ref{Fig:Anti-PT}~(a). The effective Hamiltonian for the magnon modes is 

\begin{align}
    \mathcal{H}_{\rm eff} &= \Omega \hat{m}^\dagger_1\hat{m}_1 - \Omega  \hat{m}^\dagger_2\hat{m}_2
    - i(\gamma + \Gamma)(\hat{m}^\dagger_1\hat{m}_1 + \hat{m}^\dagger_2\hat{m}_2)  \nonumber \\ &-i\Gamma (\hat{m}^\dagger_1\hat{m}_2 + \hat{m}_1\hat{m}^\dagger_2)
    \label{eqn:Anti-PT}
\end{align}
where $\Omega = (\Omega_1 - \Omega_2)/2$ is the detuning between magnon modes and $\Gamma >0$ is the strength of the effective magnon-magnon coupling, where $\Gamma$ is a real number. In this set-up, the anti-$\PT$ phase transition is observed when the eigenvalues go from purely imaginary to having a non-zero real and imaginary part, which can be observed in cavity reflection spectra as shown in Fig.~\ref{Fig:Anti-PT}~(b-c).

These are just a few examples of the abundance of non-Hermitian Hamiltonians that can be engineered in cavity magnonics. An intriguing next question then is whether these systems can be used for applications, beyond studying the spectral properties of the non-Hermitian Hamiltonians themselves. To that end, multiple proposals point toward the intriguing possibility of robust entanglement generation in the presence of dissipation or even because of it~\cite{Yuan2020, Yang2021, Ren2022, Han2022}. For example, Ref.~[\onlinecite{Yuan2020}] showed theoretically it is possible to generate a magnon-photon Bell state in a $\PT$-symmetric cavity magnon system. Somewhat counter-intuitively, when the system is in the $\PT$-broken phase, one can generate more stable entanglement than in the $\PT$-unbroken phase where the eigenvalues are entirely real. Dissipation stabilized tripartite entanglement between two photons and one magnon was theoretically studied in Ref.~[\onlinecite{Han2022}], where they reached a similar conclusion. These results can be understood in the context of reservoir engineering~\cite{Poyatos1996}, where a carefully engineered environment can enhance coherence in an open quantum system rather than suppressing it. More experimental studies are needed to show if the quantum limit can be reached such that the magnon-photon entanglement is robust and long-lived.

In addition to entanglement generation, cavity magnonics is also well positioned to study non-reciprocity in non-Hermitian systems. Non-reciprocity can occur in multimode systems where magnons couple selectively to some microwave photon modes and not others, depending for example on the photon polarization~\cite{Zhang2020}. Another definition of non-reciprocity is where the coupling between modes is assymmetric, for example in Eq.~\eqref{Eqn:HybridHamiltonian} if $|g| \neq |\tilde{g}|$. This is another type of non-Hermitian Hamiltonian which could hold important applications for signal generation~\cite{Zhang2020} and sensing~\cite{Lau2018} and merits further study. It may be particularly fruitful for applications to consider the effects of non-reciprocity in multimode systems where $M + N >2$. Yu \textit{et al.} present an intriguing proposal in this direction which would use an array of YIG spheres coupled to a single microwave waveguide to realize chiral magnon-photon coupling~\cite{Yu2020magnon, Yu2020chiral}.

Here we briefly remark on the prospects of exploring magnon-phonon coupling in non-Hermitian systems. Magnon-phonon coupling can be explored in so-called `magnomechanical' systems, where the magnons couple to mechanical deformations of the medium (i.e. stretching or compression modes of a YIG sphere, for example). There are several recent proposals for investigating magnon-phonon entanglement in cavity magnonic systems~\cite{Li2018, Li2019, Li2019b, Wang2020}. This provides another intriguing route to realize multimode non-Hermitian Hamiltonians, because the cavity photons, magnons, and phonons can all be accessed in the system. Current magnomechanical system investigations primarily use a Hermitian formalism, while a few studies of $\PT$-symmetric magnomechanical systems~\cite{Huai2019, Wang2020} have emerged to date. However, non-Hermitian effects such as dissipative phonon-magnon coupling could also be explored, expanding the symmetries of non-Hermitian Hamiltonians that can be studied. 

Cavity magnonics can also be used to study magnons coupled to superconducting qubits. In these experiments, the magnetic  material and the qubit are placed inside a cavity, where they interact via exchange of virtual microwave photons. These interactions can be coherent, in the strongly coupled regime~\cite{Tabuchi2015, Lachance2020}, or dissipative, in the dispersive regime~\cite{Wolski2020}. This tunable platform offers another way to explore non-Hermitian effects in hybrid systems through dissipatively coupled modes. In the dispersive regime in particular, Wolski \emph{et al} showed that the coherence of the qubit depends on the average magnon population, thus the qubit itself acts as a highly sensitive magnon sensor. Other proposals for exploiting the magnon-qubit interaction in superconducting qubits include manipulating single magnons through magnon blockade~\cite{Liu2019b, Wang2020, Xie2020, Wu2021, Li2021}, where the magnon-qubit interaction inhibits multiple magnon excitations. Exploring a non-Hermitian description of the magnon-superconducting qubit interaction,  beyond the inclusion of damping terms,  could yield greater insights into qubit decoherence in a magnonic bath, as well as the effect of dissipative interactions on single-magnon manipulation. For example, considering that the magnon-qubit coupling itself can be intrinsically dissipative, this may be another way to realize anti-$\PT$-symmetry in a cavity magnonic system, without relying on adiabatic elimination of the cavity mode.

Another hybrid system coupling magnonic excitations to qubits has been discussed in the context of using magnons as a medium to manipulate distantly separated spin qubits~\cite{Trifunovic2012, Trifunovic2013, Flebus2018, Flebus2019, Muhlherr2019, Neuman2020, Candido2020, Fukami2021, Zou2021}. The spin qubits could be constructed from a number of different physical systems such as silicon quantum dots, single phosphorous atoms in silicon, or nitrogen-vacany (NV) centers in diamond. This is distinct from cavity magnonics because here the magnons are a reservoir of quantum information, whereas in the former the microwave photons play this role. The spin of the qubit interacts with the local spin density of the magnonic medium via spin exchange or through dipole-dipole interactions. Most works have focused on a Hermitian Hamiltonian description of this system, where the interaction between magnons and qubits is assumed to be fully coherent. However, recently Zou \emph{et al} explored the possiblity of a \emph{dissipative} coupling between magnons and spin qubits, and showed that the two qubits could be robustly entangled in a Bell state even in the presence a purely dissipative magnon bath. More investigations into the validity of the non-Hermitian description for magnon-spin qubit coupling could provide additional avenues for reservoir engineering the magnon reservoir in order to achieve faster and more robust entangling operations between spin qubits.

\section{Conclusion and outlook \label{Sec:Outlook}}

In this perspective, we have shown how the dynamics of magnonic and hybrid magnonic systems can be mapped onto non-Hermitian Hamiltonians through several examples. In many instances, non-Hermitian theories bring to light phenomena that are not apparent within the conventional theoretical descriptions of magnetization dynamics. A prime example is exceptional points, which have been attracting a lot of attention for their potential in sensor applications and, recently, have been shown to be connected to phase transitions of the Landau-Lifshitz dynamics. Second- and third-order exceptional points have been discovered in several purely magnonic and hybrid magnonic structures. Many-body exceptional points, which display anomalous critical fluctuations~\cite{Littlewood2020}, have been predicted in driven-dissipative Bose-Einstein condensates (BECs) in a double-well potential. Magnon Bose-Einstein condensates created in yttrium-iron garnet (YIG) films via parametric pumping~\cite{demokritov2006bose} might represent a suitable platform to explore the critical dynamical phenomena driven by many-body EPs. 

It is now up to future experimental work to probe not only the presence of magnetic EPs, but also their sensitivity to perturbations, their interdependence with the onset of phase transitions in non-linear spin dynamics, and their influence on properties that can be routinely probed in spintronic setups. Magnetic heterostructures and lower-dimensional magnets might provide an ideal platform for engineering these unique non-Hermitian degeneracies. In particular, synthetic antiferromagnets appear promising as their fabrication allows for a high degree of tunability of the parameters controlling the magnetization dynamics. The strength with which the metallic spacer layers couple (antiferromagnetically) adjacent ferromagnetic films can mediate spin pumping, i.e., non-local dissipation between the layers macro-spins, can also be tuned. Furthermore, the low dimensionality of thin-film-based synthetic AFMs allows for feasible device integration. 

Magnetic metamaterials, such as synthetic AFMs and STO arrays, might be also engineered to host non-Hermitian topological magnonic phases. The topological magnon ``lasing" edge state unveiled in Ref.~[\onlinecite{Flebus2020}]  would yield a clear experimental signature, i.e., topologically-protected microwave emission at an isolated site of the driven magnetic heterostructure. To our knowledge, such measurement would constitute the first dynamical observation of a topological magnetic phase. 

Hybrid cavity magnonics systems are well-positioned to explore some of the novel phenomena discussed here such as higher-order exceptional points. These may be realized in a multimode cavity magnonics experiment for example with two YIG spheres coupled to two or more cavity modes. Further exploration of regimes of dissipative coupling between magnons and cavity photons, or magnons and superconducting qubits, could lend greater insight into reservoir engineering in hybrid systems. Here the reservoir can provide the non-Hermitian terms that drive the hybrid magnon system toward a desired steady state, such as a maximally entangled state. Furthermore, one could study how quantum fluctuations, e.g. from the input cavity ports, affect the non-Hermitian physics. This could be particularly important for understanding the limitations of proposed non-Hermitian sensors. 

From a  theoretical standpoint, future research should address feasible realizations of non-Hermitian magnonic topological phases in higher dimensions, characterized by topological edge states with finite group velocity. In Ref.~[\onlinecite{Flebus2020}]  $\mathcal{PT}$-symmetry, broken only at the system's boundaries, is essential for isolating the edge state's dynamics from the bulk ones. In higher dimensions, $\mathcal{PT}$-symmetry appears to be invalidated in the presence of topologically nontrivial spin interactions, e.g., DMI. While this might strike as a significant setback, we shall remember that the field of non-Hermitian magnetic topological phases is just emerging, and thus, there might be several alternative routes that have not been explored yet.

As we mentioned at the beginning,  non-Hermitian Hamiltonians are an approximation to more complicated dissipative dynamics, and thus, their validity is limited to certain regimes. The vast majority of non-Hermitian Hamiltonians discussed in this perspective derive from the linearization of the LLG dynamics and can give valuable insights into its dynamical and topological phase transitions. However, things become far more complex when one strives to describe the microscopic magnetization dynamics, particularly its dissipative nature away from the long-wavelength limit. To our knowledge, in this regime, there is not yet an established theoretical framework for dissipative magnetization dynamics from which one can derive the corresponding non-Hermitian Hamiltonian self-consistently.

It is well known that dissipation might be chiral and non-local in quantum hybrid systems comprised of, e.g., spin qubits coupled via a given bath. This suggests that magnon dynamics might display similar features when considering spins in contact with the baths, e.g., phononic and electronic, that a crystalline environment provides.  If one includes such terms phenomenologically in, e.g., a lattice model, the later can engender exotic phenomena, such as the skin effect, as shown by Ref.~[\onlinecite{SkinMag}]. However,  a phenomenological classical approach conceals the microscopic mechanisms that might underlie the skin effect and the regimes and times over which it takes place. These considerations suggest the need for a comprehensive theory of the magnetization dynamics and its dissipation. Such theory could be developed within the framework of the full master equation, which would include a microscopic description of the baths. In this framework, the dynamics of the collective magnetic fluctuations naturally obey an effective non-Hermitian Hamiltonian while also being subjected to Lindbladian evolution. A systematic study of the regimes and limits in which such comprehensive theory approaches the classical Landau-Lifshitz-Gilbert dynamics could shed light on several phenomena and would be a stepping stone in the theory of magnetization dynamics.

% If in two-column mode, this environment will change to single-column format so that long equations can be displayed. 
% Use only when necessary.
%\begin{widetext}
%$$\mbox{put long equation here}$$
%\end{widetext}

% Figures should be put into the text as floats. 
% Use the graphics or graphicx packages (distributed with LaTeX2e).
% See the LaTeX Graphics Companion by Michel Goosens, Sebastian Rahtz, and Frank Mittelbach for examples. 
%
% Here is an example of the general form of a figure:
% Fill in the caption in the braces of the \caption{} command. 
% Put the label that you will use with \ref{} command in the braces of the \label{} command.
%
% \begin{figure}
% \includegraphics{}%
% \caption{\label{}}%
% \end{figure}

% Tables may be be put in the text as floats.
% Here is an example of the general form of a table:
% Fill in the caption in the braces of the \caption{} command. Put the label
% that you will use with \ref{} command in the braces of the \label{} command.
% Insert the column specifiers (l, r, c, d, etc.) in the empty braces of the
% \begin{tabular}{} command.
%
% \begin{table}
% \caption{\label{} }
% \begin{tabular}{}
% \end{tabular}
% \end{table}

% If you have acknowledgments, this puts in the proper section head.
\begin{acknowledgments}
HMH acknowledges support from the San Jos\'{e} State University Research, Scholarship, and Creative Activity assigned time program. BF was supported by the NSF under Grant No. NSF DMR-2144086. 
\end{acknowledgments}

\section*{Author Declarations}

\textbf{Conflict of Interest.} The authors have no conflicts to disclose.
\textbf{Data Sharing.} Data sharing is not applicable to this article as no new data were created or analyzed in this study.

% Create the reference section using BibTeX:
\bibliography{main}

\end{document}